\documentstyle[12pt]{article}
\begin{document}
\vspace{2cm}
\centerline{\bf P.G.Grinevich, S.P.Novikov}
\vspace{1cm}
\centerline{\bf Nonselfintersecting magnetic orbits on the plane.}
\centerline{\bf Proof of Principle of the Overthrowing of the Cycles.{
\footnote{This paper will be published in 1995 by American Mathematical
Society in proccedings of the seminar by S.P.Novikov in seria ``Advances in
Mathematical Sciences''.}}}

\vspace{2cm}

{\bf 1.Introduction. Overthrowing of the Cycles. Unsolved problems}
\vspace{1cm}

Beginning from 1981 one of the present authors (S.Novikov)
published a series of papers
\cite{nov1,nov2,nov3,nt} (some of them  in collaboration with
I.Schmelzer and I.Taimanov) dedicated to the development of the analog
of Morse theory for the closed 1-forms--multivalued functions and functionals
--on the finite- and infinite-dimensional manifolds
({\bf Morse-Novikov Theory}). This theory was developed very far
for the finite dimensional manifolds (many people worked in
this direction later). The notion of {\bf ''Multivalued action''} was
understood
and {\bf ''Topological quantization of the coupling constant''} for them
was formulated by Novikov in 1981 as a Corollary from the requirement,
that the {\bf Feinmann Amplitude should be one-valued on the space
of fields--maps}, by  Deser-Jackiv-Templeton in 1982
for the special case of Chern-Simons functional and by Witten in 1983).
This idea found very important applications in the quantum field theory.
Very beautiful
analog of this theory appeared also in the late 80-ies in the Symplectic
Geometry and Topology, when the so-called Floer Homology
Theory was discovered.

A very first topological idea of this theory, formulated
in early 80-ies, was the so-called {\bf Principle of the Overthrowing
of the Cycles. It led to the results which were not proved
rigorously   until now}.  Our goal is to prove  some of them.

We remind here that Novikov studied
in particularly an important class of classical Hamiltonian
systems of the different
physical origin, formally equivalent to the motion of the charged
particle
on the Riemannian manifolds $M^n$ in the external magnetic field
$\Omega$, which is a closed 2-form on the manifold (see \cite{nov3}).
In terms of Symplectic Geometry, these Hamiltonian Systems
on the Phase spaces like $W^{2n}=T^{\ast}(M^n)$
are generated by the standard
Hamiltonian functions (the same as the so-called ''Natural Systems''
in Classical Mechanics) corresponding to the nonstandard Symplectic
Structure, determined by the External Magnetic Field.
In the most interesting cases
our symplectic form is topologically nontrivial (i.e. it may have
nontrivial
cohomology class in $H^2(W^{2n},R)$).

Periodic orbits are the
extremals of the (possibly multivalued) action functional $S$
on the space $L(M^n)$ of the
closed loops (i.e. smooth or piecewise smooth
mappings of the circle in the manifold
$M^n$):
$$S\{\gamma(t)\}=\oint_{\gamma}1/2\left(\frac{d\gamma}{dt}\right)^2+
e\oint_{\gamma}d^{-1}(\Omega)$$

This quantity is not well-defined in general as a functional,
but its {\bf variation $\delta S$ is  well-defined
as a closed 1-form on the space of
closed loops $L(M^n)$} (this is the {\bf situation of Dirac monopole}).

Even if the closed 1-form $\delta S$ is exact,
its integral $S$ may be not bounded from below. In both these cases
 standard Morse theory
does not work. For the fixed energy $E$ we  replace the action
functional by the {\bf ''Maupertui--Fermat''  functional}
with the same extremals:
$$F_E(\gamma)=(2E)^{1/2}l(\gamma)+e\oint_{\gamma}d^{-1}\Omega$$

This functional is also multivalued in general. Here $l(\gamma)$
is an ordinary Riemannian
length. Let the charge $e$ will be equal to 1 and the form $\Omega$
is exact $\Omega=dA$ and small enough. The functional above is positive.
We have some very nice special case of the Finsler metric (its
 geometry was investigated by E.Cartan many years ago). We may apply the
 ordinary Morse-Lusternik-Schnirelmann theory in this case.

\newtheorem{itr}{Definition}
\begin{itr}
 We call the functional $F_E$ {\bf not everywhere positive} if
 the form $p^{\ast}\Omega$ is exact on the universal covering
 $p:M\rightarrow M^n$ and there exist
 a closed curve $\gamma$ on the universal covering (or the curve
 homotopic to zero in $M^n$),
 such that $F_E(\gamma)<0$. By definition $F_E(\gamma_0)=0$ for any
 constant curve $\gamma_0$.
 \end{itr}

 \begin{itr}
We call the functional $F_E$ {\bf essentially multivalued}
if the form $p^{\ast}
\Omega$ is not exact on the universal covering $M$. It is well-defined
as a functional on some regular nontrivial free abelian
covering space $\hat{L}\rightarrow L(M)$ with
 discreet fiber $Z^k$ over the loop space for the manifold $M$.
\end{itr}

In this case there is
a natural imbedding of the (trivial) covering space over the one-point curves
$M\times Z^k\subset \hat{L}$, such that $F_E(M\times 0)=0$
for some selected point $0\in Z^k$ and $F_E(M\times j)\neq 0$ for $j\neq 0$.

In the last case the functional $F_E$ is obviously
not everywhere positive on the
covering space $\hat{L}$. There exist an index $j\in Z^k$ such that
$$F_E(M\times j)<0$$

There is a natural free action of the group $\pi_1(M^n)$ on the loop space
$L(M)$, such that the factor space is isomorphic to the space
$L_0(M^n)$. Here $L_0(X)$
denotes the space of loops on $X$, homotopic to zero.
This action extends naturally to the space $\hat{L}$ and
we are coming to the factor-space $\hat{L}_0$ of the space $\hat{L}$ by
the group $\pi_1(M^n))$.

On the last space $\hat{L}_0$ our functional $F_E$ is well-defined
and not everywhere positive. There is an imbedding
$$M^n\times Z^k\subset \hat{L}_0$$

such that
$$F_E(M^n\times 0)=0,F_E(M^n\times j)<0$$

for the same indices as above. It makes sense only if our functional is
essentially multivalued: there exist an index $j$ different from zero.
For one valued  functionals we have $k=0$
and $Z^k,k=0$ contains only one point $0$.

The following two lemmas are trivial, but important.

\newtheorem{lem}{Lemma}
\begin{lem}
 All imbeddings
$M^n\times j\rightarrow \hat{L}_0$ are homotopic to one with index 0.
\end{lem}

\begin{lem}
Our functional has nondegenerate manifolds of local minima in
all one-point families $M_j=M^n\times j\subset \hat{L}_0$.
\end{lem}

We are coming now to the following important definition.

\begin{itr}
By the {\bf Overthrowing of the Cycle} (set) $Z\subset M^n$
(in the negative domain)
for the given multivalued or
not everywhere positive functional $F_E$ we call any continuous map
$$f:Z\times I[0,1]\rightarrow \hat{L}_0$$

such that $f(Z\times 0)=Z\subset M_0$ and  $F_E(Z\times 1)<0$.
\end{itr}

The existence of such overthrowing was pointed out in early
80-ies by Novikov as a main topological reason
for the existence of periodic orbits, homotopic to zero,
in the magnetic field. There are two important examples.

\newtheorem{exm}{Example}
\begin{exm}
For the essentially multivalued
functionals we may take $Z=M^n$.
Overthrowing here is a homotopy between $M^n\times 0$ and
$M^n\times j$ as above.
\end{exm}

\begin{exm}
 For the case of one-valued but
not everywhere positive functionals we may take $Z$ as one point
in $M^n$. Later Taimanov proved in \cite{tai1} that there exist
overthrowing with $Z=M^n$ for any not everywhere positive functional.
\end{exm}

As a Corollary from the overthrowing an {\bf analog of the Morse
inequalities} was formulated. Let all critical points are nondegenerate.
 For the number of them with Morse index equal to $i$ and {\bf positive value
 of the functional} we have inequality:
$$m_i(F_E)\geq b_{i-1}(M^n),i\geq 1$$

Here $b_i$ are Betti numbers or any their improvements of
the Smale type. Critical points may be degenerate or they may
be multiples of one smaller extremal. Therefore we expect to prove
existence of
 one periodic extremal from this arguments. However
there exist an important difficulty (pointed out  by
Bolotin many years ago):

{\bf We  prove the existence of the
positive critical values $c_s>0$ for the functional $F_E$ by the minimax
arguments, but actual
critical points may not exist. Our functional violates the important
 Compactness Principle.}

 The critical value $F_E=c_s>0$ may be realized by the infinitely long curve
 $\gamma$,
 which satisfies to the Euler-Lagrange equation and may be approximated by
 the locally convergent sequence of the closed curves
 $\gamma_i(t)\rightarrow \gamma(t)$,
 such that:
 $$F_E(\gamma_i)\rightarrow c_s+0,l(\gamma_i)\rightarrow\infty$$

 Until now we don't know any examples of such infinitely long extremals
 obtained through the overthrowing of the cycles.

 {\bf
 In the present paper we
 are going to prove completely the Overthrowing Principle for the
 important case $M^2=T^2$ with Euclidean metric and arbitrary
 nonzero magnetic field.}

 We may think about the
 Euclidean plane $R^2$ with everywhere nonzero double-periodic
 magnetic field, directed along
 $z$--axis. {\bf All  4 periodic extremals for any generic energy $E$
 (with the
 Morse indices 1,2,2,3 of the Maupertui-Fermat functional) will be found as
 convex nonselfintersecting curves.
 Therefore they are geometrically distinct.} Of course
 we obtain other extremals from them by the discreet translations on periods.
 In principle, homological arguments don't give anything else.

 {\bf Remark 1.} In the paper \cite{nt} after long story
 a right criteria  were found (in the Theorem 1)
 for the existence of the nonselfintersecting
 extremals of multivalued and not everywhere
 positive functionals on the 2-sphere. The idea of the proof was incomplete
 for the essentially multivalued case. Later it was completed
 and finally proved by  Taimanov (see proof and all history
 of this problem in the survey article \cite{tai2}).

 {\bf Remark 2.} It is clear for us now that no analogs
 of the Morse type theory can be
 constructed   on the space of immersions. Therefore the theorem 2
 of the paper \cite{nt} is unnatural.Its most general  form
 is probably  wrong. It should be replaced by the
 stronger result--by the main theorem of
 the present paper for the nonzero magnetic field.

{\bf Remark 3.} In the very interesting papers \cite{gin1,gin2}  V.Ginzburg
proved the existence of periodic orbits with energy small enough and large
enough, using the perturbation of the limiting pictures.
In particularly, he pointed out to us that the theorem 3, announced
without proof in the paper
\cite{nt}, is wrong. In fact, it contradicts to the example
of the constant negative curvature and magnetic field equal
to the Gaussian 2-form with the right sign, such that the extremals
are exactly the horocycles: there is no periodic horocycles on the
compact surfaces\footnote{As Ginzburg wrote, this example
was pointed out to him by Marina Ratner}.This mistake is interesting:
theorem 3 was extracted from
the lemma 3 which claims that our functional is bounded
from below in any free
homotopy class of loops, if it is true for the trivial one. This lemma
is  wrong. It is true
for the homotopy classes of mappings $(S^1,s)\rightarrow (M^2,x)$
representing
any element of fundamental group $\pi_1(M^2,x)$, but may be wrong for some
free homotopy classes containing the infinite number of elements of
fundamental group. It is exactly what is going on in this counterexample.

For the finite conjugacy classes our theorem can be true.
However there is no proof of it: the lack of compactness
presents here analogous difficulty as before.

We may find some finite critical value $c>-\infty$,
but corresponding
extremal may be infinitely long as before. Which kind of extremal we may
get? For the surfaces with negative curvature and horocycles
we have $c=-\infty$, so this case is out of our arguments.

Nontrivial example we get on the 2-torus $T^2$ with the
exact magnetic field $\Omega=dA$, such that our functional $F_E$
is positive on the space of the closed curves homotopic to zero.

This property is always true for the energy, larger than some
critical energy $E_0$. {\bf In many cases an interval of  energies
exists such that the
Maupertui-Fermat functional  is not a well-defined Finsler metric,
but positive
on the space of loops homotopic to zero.}

Another interesting example we get for $n=3$. Let the manifold $M^n$
is a fiber bundle with fiber isomorphic to the circle $S^1\subset M^3$.
For the magnetic fields $\Omega$ with homology classes from the base
we may ask about the periodic extremals homotopic to the fiber.
For the 3-torus this restriction means that our magnetic field
has no more than one rationally independent flux over the integral
2-cycles.

\pagebreak

\vspace{2cm}
\centerline{\bf 2.Nonzero double periodic magnetic field on the plane. }
\centerline{\bf Proof of Overthrowing of the Cycles for Convex Polygons.}

\vspace{1cm}
We consider now nonzero smooth double periodic magnetic field on the
Euclidean plane,
directed orthogonal to this  plane $R^2$:
$$B(x+1,y)=B(x,y+1)=B(x,y)>0$$

For the energy level such that
$$(2E)^{1/2}=1$$

we denote a Maupertui-Fermat functional
by $F=F_E$. We shall consider only this case without any
losses of generality.

Consider the space of closed convex curves, oriented
in such a way that:
$$F\{\gamma\}=l(\gamma)-\int\int_{K}B(x,y)dxdy,x^1=x,x^2=y$$

In this formula $K$ means a positively oriented domain inside
of the curve $\gamma$,
magnetic field $B$ is positive. The second term we
call a {\bf magnetic area}. It comes with the negative sign.

In this section we consider the functional $F$ on the {\bf space $P_N$
of the
straight-line convex ''parameterized''
polygons $\gamma\in P_N$ containing exactly $N$
equal straight-line pieces of any length $L$}. By definition,
''parameterized polygon'' means   ''polygon with
some natural numeration of vertices''
$$AB...CDA=A_1A_2...A_{N-1}A_NA_{N+1}
=A_1$$

Cyclic permutation of this numeration leads to the free action
of the group $Z/N$ on the space $P_N$. We denote factor-space by
$\bar{P}_N$.

Let $B_{min},B_{max}$ denote minimum and maximum of $B(x,y)$
on the torus $T^2$. We introduce the following parameters:
$$N_0=\left[\frac{8B_{max}}{B_{min}}\right]+1$$
$$\alpha_0=\min\{\frac{1}{1000N},\frac{2}{N}\arctan\left(\frac
{9B_{min}}{20B_{max}(2N^3+N/2)}\right)\}$$
$$L_0=\frac{4N}{\sin\left(\frac{
\alpha_0}{2}\right)B_{min}}$$

This parameters depend from $N,B_{min},B_{max}$.

We shall consider the spaces $P_N$ for $N>N_0$ only. For $N\rightarrow
\infty$ we have $\alpha_0\rightarrow 0$
and $L_O\rightarrow\infty$. Let $AB,BC$ are the neighboring edges
of the convex polygon. In the point $B$ we have an {\bf external angle}
$\alpha$ and {\bf internal angle} $\beta$, such that $\alpha+\beta=\pi$.

\begin{itr}
We call a convex closed polygon from the space $P_N$
{\bf admissible} if all its external angles are larger than
$\frac{\alpha_0}{2}$ and $L<2L_0$
\end{itr}

On the subspace of admissible polygons $P^a_N$ we define a {\bf corrected
functional}:
$$F_a(\gamma)=F(\gamma)+\sum_{k=1}^N\phi(\frac{\alpha_0}{\alpha_k})+
L_0\psi(\frac{L}{L_0})$$

Here $\alpha_k$ means the k-th external angle of our admissible
convex polygon $\gamma$, $\phi$ and $\psi$ are such real nonnegative
functions on the closed interval  $[0,2]$ that
$$\phi(x)=\psi(x)=0, x\leq 1$$
$$\frac{d\phi}{dx}>0,\frac{d\psi}{dx}>0, x>1$$
$$\phi(x)\rightarrow+\infty,\psi(x)\rightarrow 1,x\rightarrow 2$$

and both $x$-derivatives of this functions converge to the
$+\infty$ if $x\rightarrow 2$.

\newtheorem{theo}{Theorem}
\begin{theo}
Let $\gamma\in P_N$ is a convex polygon with $N>N_0$. It is an extremal
for the functional $F$ on this space such that $F(\gamma)>0$
if and only if
$\gamma$ is an admissible curve, an extremal also for  the functional $F_a$
and $F(\gamma)=F_a(\gamma)>0$.
\end{theo}

The obvious geometric facts are true:

\begin{lem}
Let $\gamma_i, i=1,2$ are two convex polygons, such that
$\gamma_1$ lies completely inside of $\gamma_2$. In this case
we have $l(\gamma_1)<l(\gamma_2)$.
\end{lem}

\begin{lem}
Let $\gamma\in P_N$ is a convex polygon with total length $NL$,
such that there exist
two internal angles in it less than $\frac{\pi}{3}$. It follows
that the distance between these two vertices (say $A,B$)
is at least $LN/4$.
\end{lem}

For the proof of last lemma we observe that this vertices
$A,B$ cannot be neighboring in $\gamma$. All our curve
$\gamma$ belongs to the interior of  the romb
 whose two opposite vertices are
exactly $A,B$ with corresponding internal angles equal to $\frac{2\pi}{3}$
for this romb. The perimeter of the romb is less than $4|AB|$.
We conclude therefore that $LN<4|AB|$ from lemma 3.
Lemma is proved.

\begin{lem}
Let $D$ is any convex subset in the Euclidean space $R^2$, bounded by
the polygon $\gamma\in P_N,l(\gamma)=NL,$ and $C$ is any point inside it.
After the rotation of this set on the small angle $\delta\alpha$
around the point $C$ we get a domain $D_1$ whose magnetic area
satisfies to inequality:
$$|\int\int_{D_1}Bdxdy-\int\int_DBdxdy|<N^2L^2(B_{max}-B_{min})\frac{\delta
\alpha}{2}$$

\end{lem}

For the proof of this lemma we point out that after the rotation on the
small angle $\delta\alpha$ total set $D_1$ minus the original one will have
area no more than $N^2L^2\frac{\delta\alpha}{2}$. Combining this with
the obvious estimates for the integral we get our final estimate.
Lemma is proved.

\begin{lem}

Let $\gamma\in P_N,N>N_0$ and there exist two different vertices
(say, $A,B$), such that
the corresponding internal angles     are less than $\frac{\pi}{3}$.
There exist
a small deformation  $\gamma_t$ of the curve $\gamma=\gamma_0$
in the space $P_N$ (with fixed length),
such that magnetic area increases in the linear approximation and all
external angles don't decrease. Therefore the curve $\gamma$ cannot be an
extremal for any one of the functionals $F,F_a$.
\end{lem}

Proof of the lemma. Let the segment $AB$ be horizontal in our picture.
It divides $\gamma$ on 2 pieces $AC\ldots DB$ (upper piece)
 and $AE\ldots FB$ (lower piece). Our deformation will be such that
 the lower piece $E\ldots F$ does not move $E_t=E,\dots F_t=F$
 and the upper piece
 $C\dots D$ moves up perpendicular to the segment $AB$ parallel to itself on
 the distance $t$, $(C_t\ldots D_t)||(C\ldots D)$.

 Position of the vertices $A_t,B_t$ we define from this completely,
 because the length $L$ does not change.

 This
 deformation has the desired properties (see elementary
 trigonometric calculation in the Appendix). Lemma is proved.

So we cannot have two internal angles less than $\frac{\pi}{3}$
for the extremals.

\begin{lem}
No polygon $\gamma\in P_N$ can have all  external
angles  except may be one angle (say, $\alpha_N$) less than
$\alpha_0$.
\end{lem}

Proof of this statement follows immediately from the definition of
$\alpha_0$ and elementary geometric facts: total sum of all
external angles is equal to $2\pi$, each of them is less than $\pi$.
Therefore we have
$$\sum_{k=1}^{N-1}\alpha_k=2\pi-\alpha_N< (N-1)\alpha_0\leq
\frac{N-1}{1000N}$$

At the same time we have $\alpha_N<\pi$. It leads to the contradiction,
which proves our lemma.

Consider now a curve (polygon from the space $P_N$) containing at least
two vertices with external angles more than $\alpha_0$. Let this vertices
are  $A,B$ and all vertices between them (from one side)
have ''small''  external angles (i.e.
less than $\alpha_0$). We construct a deformation  $\gamma_t$ of this curve
$\gamma=\gamma_0$ in the space $P_N$ with fixed length:
let $[AB]$ is a segment between this two points and $C,D$ are the
vertices with orthogonal projections on the segment $[AB]$ closest
to the centrum. Here $C$ belongs to the arc with small external angles
and $D$ belongs to the other arc of $\gamma$. Let $A_t=A$ and
$B_t$ is obtained from $B$ by the small shift $\delta x=t$
along the segment $[AB]$
in the direction of $A$. We rotate the arcs $A\ldots C$ and $A\ldots D$
around the point $A=A_t$. The arcs $B\ldots C$ and $B\ldots D$ we shift
parallel to themselves on the same distance as the point $B$. After that we
rotate them around the point $B_t$. Finally we  find the points
$C_t,D_t$ as a crossing points. Following lemma is true for this deformation.

\begin{lem}
The deformation $\gamma_t$ described above does not change the length.
It is such that
all external angles (except the angles in
the vertices $A,B$ with external angles more
than $\alpha_0$)  don't decrease; $t$-derivative of the magnetic area
for $t=0$ is nonzero. Therefore the curve $\gamma$ cannot be an extremal
for the functionals $F,F_a$; {\bf  Any curve which is an
extremal for each of them
is such that all
external angles are more than $\alpha_0$}.
\end{lem}

Proof of this lemma uses the lemmas above. It is based on the
elementary trigonometric calculations using the values of
parameters $\alpha_0, N_0, L_0$, fixed in the beginning of this section
(see Appendix for the details).

\begin{lem}
Let the curve $\gamma\in P_N$ is such that all external angles
are bigger than $\alpha_0$ and $L>L_0$. In this case we have
$F(\gamma)<0$. If the curve is admissible $L<2L_0$ we have $F_a(\gamma)
<0$
\end{lem}

For the proof of this lemma it is enough to estimate the
   magnetic area of any triangle $ABC$ based on two edges $AB,BA$
   of our polygon. The external angle is bigger than $\alpha_0$.
   Therefore its area $S$ is bigger than $S_0$:
   $$S>S_0=(1/2)L^2 \sin(\frac{\alpha_0}{2})$$

   and its magnetic area is bigger than  $B_{min}S_0>2NL$. We have
   $l(\gamma)=NL$. As a corollary we are coming to inequality:
   $$F(\gamma)<NL-2NL<0$$

   All external angles are bigger than $\alpha_0$. So the contribution of the
function
   $\phi$ in the value of the functional $F_a(\gamma)$ is equal to zero.
   By definition, we always have $\psi\leq 1$. Therefore we conclude
   for $L>L_0$ that
   $$F_a(\gamma)=F(\gamma)+L_0\psi(\frac{L}{L_0})<-NL+L_0<0$$

Lemma is proved.

Proof of the theorem 1 follows now from the lemmas.
{\bf Theorem 1 is proved}.

We are going to construct now a natural analog of the
Morse theory for the functional $F_a$ on the space $P_N^a$ of the
admissible polygons--or, more exactly on the space $\bar{P}_N^a$
of the admissible polygons completed naturally
by the one-point curves and
factorized by the discreet group $Z^2\times(Z/N)$
generated by the basic translations
of the plane $R^2$ and cyclic permutation of the order of  vertices.

 This space is homotopy equivalent to the torus $T^2$
(i.e. to the subspace of the one point curves). This space without
one point curves is homotopy equivalent to the 3-torus $T^2\times S^1$.

 We are going to use the Morse type estimates
''modulo subspace $P^0$'',
where the functional $F_a$ is less or equal to zero
$$P^0=\{F_a\leq 0\}$$

An easy lemma is true:

\begin{lem}
The space $P^0$ is not connected. It contains at least two components.
One of them is exactly set of all one point curves $T^2_0\subset P^0$.
Another one $P^0_1$ contains all $N$-polygons $\gamma$
with equal angles, such that
the length of edges $L$ is big enough (but less than $2L_0$).
\end{lem}

We already know that the set of one-point curves is a local minimum
 for the functional $F$. By definition, the value of the functional $F_a$
 on it is equal to zero. So this set is a local minimum also for $F_a$,
 because $F_a\geq F$ for any admissible curve. Therefore the set of one
 point curves is  isolated in $P^0$. The curves with equal angles
 have all external angles equal to $\frac{2\pi}{N}>\alpha_0$.
 For $L=L_0$ and large $N>N_0$ we have $F=F_a$ for them and $F<0$.

 Lemma is proved.

For any point in the plane we construct an Overthrowing of it,
continuously depending from this point.

\begin{itr}
 By definition, our Initial Overthrowing
is a set of all $N$-polygons with centrum in this point and
with equal angles and edges with the length $L<L_0$.
It determines a map
$$f:(T^2\times S^1)\times [01]\rightarrow \bar{P}_N^a$$

such that
$$f(T^2\times S^1\times 0)=T^2_0\subset P_0, f(T^2\times S^1\times 1)
\subset P^0_1$$
Parameter along the circle $S^1$ here is exactly an angle numerating
all polygons with the same centrum and same length,
parameter in the interval $[01]$
coincides with  radius divided by the maximal radius, such that
all image belongs to the negative values of our functionals for all
central points in the plane.
\end{itr}

An obvious lemma is true:

\begin{lem}
An Overthrowing
$$f:(T^2\times S^1\times [01],T^2\times S^1\times 0\bigcup T^2
\times S^1\times 1)\rightarrow
(\bar{P}_N^a,P^0)$$

generates monomorphisms in Homology groups:
$$H^{i-1}(T^2\times S^1)\rightarrow H^i(\bar{P}_N^a,P^0),i\geq 1$$

 \end{lem}

Our functional $F_a$ generates a cell decomposition of the
space $\bar{P}_N^a$ modulo $P_0$, corresponding to the critical points
such that $F_a(\gamma)>0$. This is a corollary from our lemmas, because
this space is invariant under the gradient flow
(all gradient lines go inside of it). So we may apply standard arguments
of the Morse theory
to this  space modulo negative subspace $P_0$. Combining this fact with
the previous lemma, we are coming to the theorem:

\begin{theo}
For any value $N>N_0$ there exist at least two different extremals
of the  Maupertui-Fermat functional $F$   in the space $P_N^a$
of the admissible convex polygons. If critical points are nondegenerate,
there exist at least 8 of them in the same space with  Morse indices
equal to 1, 2, 2, 2, 3, 3, 3, 4.
\end{theo}

Proof of the theorem. By the minimax principle, we always have
at least one extremal in this space. Let we have only one critical point.
After the long gradient deformation, starting from
the initial Overthrowing  Process
$$f=f_0:T^2\times S^1\times [01]\rightarrow \bar{P}_N^a$$

we are coming to the new Overthrowing Process $f_1$ in which
almost all image is below the critical
level
and the remaining part is concentrated in the small neighborhood
of the critical point. After removing from the space
$\bar{P}^a_N$
 some small neighborhood of the critical point the new overthrowing
will split on some pieces (at least two) such that the image of
its boundaries
$T^2\times S^1\times 0$ and $T^2\times S^1\times 1$
belongs to different components.

It follows from the fact that any new overthrowing of one point should pass
through the same small neighborhood as the new overthrowing of all
torus $T^2$. By any new overthrowing of the point we have in mind
any curve $f_1:\tau(t)\rightarrow \bar{P}_N^a$ where $\tau(t)$ is
any continuous curve
in $T^2\times S^1\times [01]$ such that $\tau(0)\in T^2\times S^1
\times 0,f_1(\tau(1))\in P^0$

Our space $P_N^a$ is locally contractible. Using this,
we deform its identity map
onto itself in such a way that after deformation all small neighborhood
of the critical point will collapse to this point.
Finally we constructed a deformation of the set of one point curves
to one (critical) point in the space $P_N$. However this
set is nonhomotopic to zero in the space $P_N$. We are coming
to contradiction.  So we  have at least two extremals.

Other part of this theorem is a standard obvious
corollary from the handle decomposition generated by the critical points of
the
functional $F_a$. It follows from the lemmas above that we may apply the
standard arguments of Morse Theory here.
Theorem 2 is proved.

\pagebreak

\vspace{1cm}

{\bf 3.Compactness Property for  $N\rightarrow \infty$. Main results.}

\vspace {1cm}

\begin{itr}
For any convex polygon $\gamma\in P_N$ we call by the
{\bf Maximal Diameter}
$D_{max}$ a maximal distance between two points of this polygon.
By the diameter in the direction $\phi$ we call a maximal
distance $D_{\phi}$
between two  straight lines parallel to the direction $\phi$,
which have nontrivial intersection with $\gamma$. We call by the
 maximal and minimal diameters $D_{max},D_{min}$ exactly maximum
 and minimum of the function $D_{\phi}$, corresponding to the directions
 $\phi_{max},\phi_{min}$.
 \end{itr}

\begin{theo}
Let $\gamma_N\in P_N$ is an extremal of the functional $F$,
such that $F(\gamma_N)>0$.Following  estimate for its maximal diameter
and for the maximal length ${\cal L}_0$
are true:
$$D_{max}\leq \frac{8(3+B_{max}B^{-1}_{min})}{B_{min}(1-8N^{-1})}, \
{\cal L}_0 \le 4 D_{max}
$$

\end{theo}

 The proof of this theorem follows from the lemma:

 \begin{lem}
For any extremal $\gamma$  of the functional $F$ on the space $P_N$, such that
$F(\gamma)>0$, the estimate is true:
$$\frac{D_{max}}{D_{min}}\leq \frac{B_{max}B_{min}^{-1}+3}{1-8N^{-1}}$$

\end{lem}

Proof of the lemma. We describe a deformation, which preserves length
of $\gamma$  and changes  the magnetic area in the linear approximation
if the inequality
is not true.  The $y$-axis is exactly direction $\phi_{min}$ in our picture.

Let $AB,HG$ are the most left edges such that their angles with $x$-axis
are no more than $\frac{\pi}{4}$ and $CD,FE$ are the
rightest edges with the same property.
The arcs
$AB\ldots CD$ and $HG\ldots FE$ belong to the
upper and lower parts of $\gamma$.
The points $A$ and $H$ or $D$ and $E$ may coincide, but it is not important.
Our deformation $\gamma_t, \gamma_0=\gamma$ is such that   $A_t\ldots H_t
=A\ldots H$,
the arc $D_t\ldots E_t$ is obtained from  $D\ldots E$   by the parallel
shift on the distance
$\delta x=t$ to the left. The arc $B_t\ldots C_t$ is obtained by the
parallel shift of the arc $B\ldots C$ up (on the distance $\delta y_1$
and left (on the distance $\delta x_1$). The arc $G_t\ldots F_t$ is obtained
from $G\ldots F$ by the parallel shift down (on the distance $\delta y_2$)
and left (on the distance $\delta x_2$.
The value of all this parameters as function from the variable $t$
follows from the requirement that
all lengths are the same and the new polygon $\gamma_t$ is closed.

Proof of this lemma follows from the trigonometric calculations
(see Appendix).

Proof of the Theorem: let $A,B,C,D$ are the most left, most upper,
rightest and most lower vertices in $\gamma$. A polygon $ABCD$
with four edges belongs completely to the interior of $\gamma$.
Therefore its area is at least $\frac{D_{min}D_{max}}{2}$
and the magnetic area $Q$ is at least
$1/2D_{min}D_{max}B_{min}$. Combining this with the trivial
estimate $l(\gamma)\leq 4D_{max}$ and $F(\gamma)>0$ we are coming
to inequality:
$$0<F=l-Q<4D_{max}-1/2D_{min}D_{max}B_{min}$$

or finally
$$D_{min}<8B_{min}^{-1}$$

Using the lemma above, we are coming to the desired inequality
for $D_{max}$.
Theorem is proved.

Consider now a sequence $\gamma_N$ of the extremals of  functional
$F$ on the spaces $P_N$ with $F>0$ and $N>N_0$. In fact we consider only
a subsequence $N_k=N_12^k$.

For the large $N$, small external
angles and bounded total length
$NL$ the polygons  $\kappa$ and $p_N(\kappa)$ are very closed.

{} From the theorem above we conclude that {\bf There exist a
subsequence $k_j\rightarrow \infty$ such that $\kappa_j=\gamma_{N_{k_j}}
\rightarrow \gamma$ where $\gamma$ is a continuous curve,}
because all family of our extremal $N_{k_j}$-polygons $\kappa_j$
with positive value
of the functional $F$ is precompact.

\begin{theo}
The limiting curve $\gamma$ is a periodic smooth extremal of
the functional $F$ with positive value of $F(\gamma)>0$.
\end{theo}

\begin{lem}
Let $\kappa_j\in P_N,N=N_{k_j}$ as above is a sequence
of ''relative extremals'' with positive
value of the functional $F$ and $\gamma$ is a limiting continuous curve.
For $N\rightarrow\infty$ all family of external angles of
the curves $\kappa_j$
converges to zero as $O(1/N)$.
\end{lem}

Let $A$ is a vertex with largest external angle $\alpha$
and $B$ is ''opposite''
vertex, such that the arc $A\ldots B$ contains exactly $N/2$ edges.
Consider following deformation $\gamma_t$ of the polygon
$\kappa_j=\gamma_0$:
$B_t=B$, all vertices except $A$ move along their own edges towards
$B$ in such a way that the distance between any two vertices
will be exactly $L-\delta L,\delta L=t$. A shift of the vertex $A$
will be completely determined by the requirement that new polygon
has equal edges with length $L-t$.

For the variation of the functional $F$ in the point $t=0$
we get inequality
(see elementary trigonometric calculation in Appendix):
$$|\delta F|>\delta L(N-\pi{\cal L}_0B_{max}-{\cal L}_0B_{max}(\sin\alpha)^
{-1})$$

At the same time we remember that $\delta F=0$ for $t=0$.

Finally we are coming to inequality:
$$\sin\alpha<(\frac{N}{{\cal L}_0B_{max}}-\pi)^{-1}$$

Therefore we proved that for the large enough
$N$ there exist such constant $c$
that $\alpha<cN^{-1}$.
Lemma is proved.

\begin{lem}
The limiting curve $\gamma$ belongs to the class $C^1$.
\end{lem}

We found already the upper estimate for the length of ''relative extremals''
in the lemma above. It is easy to find also the lower estimate for this
length. Consider any point $x$ inside $\gamma$. We apply homotety with
centrum in this point and with coefficient $1+p$.
For the variation we have
$$0=\delta F=\delta l(\gamma)-\delta \int\int_KBdxdy>pl(\gamma)-
pl(\gamma)d_{max}B_{max}$$

Here $d_{max}$ means maximal distance from the point $x$ to $\gamma$.
We deduce from this an inequality:
$$d\geq B_{max}^{-1}, NL=l(\gamma)\geq l_0=2B_{max}^{-1}$$

Consider any arc $P...Q...R...S$ on the extremal $\gamma$, containing
$n$ edges, $n\leq N/2$. By the previous lemma, the angle $\phi$ between
the lines $PQ$ and $RS$ in the point of their intersection (outside of
$\gamma$) has the order $O(nN^{-1})$:
$$\phi \leq nc/N$$

For the length of the arc $P...S$ we have $l(P...S)=nL\geq nl_0N_{-1}$.
Combining this with previous inequality,  we get:
$$\phi\geq cl(P....S)l_0^{-1}$$

For the limiting curve $N\rightarrow\infty$ we have an upper estimate
for the angle between two ''tangent'' lines  in the points $P,S$
$$\phi\geq c(B_{min},B_{max})\times l(P,S)$$

(distance along the curve).

\begin{itr}
 By the ''tangent'' line for any convex curve we call
 any straight line
which has all our curve from one side. For the vertices of convex polygons
''tangent line'' means that it has only one common point with our polygon.
\end{itr}

Our curve is convex because
it is a limit of convex curves. Lemma follows from this estimate.

The polygons from our sequence $\kappa_j$ belong
to the spaces $P_{2^{k_j}N_1}$.
We fix  numeration such that all the vertices $P_{j,s}$  with
numbers $2^{k_j }s$
converge to some points $P_s$ on the limiting curve for fixed values of $s$,
$j\rightarrow\infty, s=1,...N_1$.

For $N_1$ large enough and any two vertices $P_{0,s}=R_0, P_{0,r}=Q_0$ on this
curve we have two sequences
$$P_{j,s}=R_j\rightarrow R,
P_{j,t}=Q_j\rightarrow Q$$

\begin{lem}
Following estimate is true for the angle $\phi$ between two straight
lines, ''tangent'' to
the polygons $\kappa_j$ in the vertices $R_j, Q_j$:
$$\phi=\int_{R_j}^{Q_j}B(\kappa_j(t))dl(t)+O(\frac{1}{2^{k_j}N_1}),
j\rightarrow\infty$$

Here integral is taken along the curve $\kappa_j$ using a natural
parameter $l$.
\end{lem}

Proof. Consider the arc $TR_jU....A...VQ_jS$ where $A$ is a ''central''
vertex between $R_j$ and $Q_j$ (or one of two central vertices, if
the number of edges in the arc is odd). Let $B$ is a ''central''
vertex of the opposite arc $R_jT....B...SQ_j$ in the same sense,
$T_1,T_2$ means two ''tangent'' lines in the vertices $R_j,Q_j$
and $\phi$ means external angle in their crossing point.

We construct such deformation $\gamma_t, \gamma_0=\kappa_j$
of this extremal that:

All arc $R_jT....B...SQ_j$ doesn't move;

The points $R_j, Q_j$ we move at first inside of the curve $\gamma_0$
on the small distance $\delta s=t$ in the direction
perpendicular to the edges
$R_jU,Q_jV$. We denote new vertices by $R^1,Q^1$.
We move all edges of the arc $R_jU....VQ_j\rightarrow R^1U^1...V^1Q^1$
inside
on the distance $\delta s$ in
the directions perpendicular to each edge and construct from their pieces
a new
arc $R^1U^1......V^1Q^1$
with slightly smaller edges (not necessary equal to each other).

After that we make all edges equal by the deformation, such
that the point $B$ does not move, all vertices from the arc
$A^1...Q^1...B^1$ move along their own edges on this arc
in the direction towards $B$,
all vertices on the arc $A^1...R^1...B$ move along their own edges
on this arc towards $B$ , the vertex $A$  moves along one of two edges
(which one will be uniquely defined by the condition that we have
to get finally  equal edges).

On the first step of deformation we have an estimate for the length:
$$\delta l=\phi\delta s+O(\frac{1}{2^{k_j}N_1})$$

(from lemma 13)

and for the magnetic area:
$$\delta\int\int_KBdxdy=\int_{R_j}^{Q_j}Bdl+O(\frac{1}{2^{k_j}N_1})$$

{}From the same lemma 13  we conclude that the total product  over all vertices
is equal to one plus something small:
$$\cos\alpha_1\times\ldots \cos\alpha_{2^{k_j}N_1}=1+O(\frac{1}{2^{k_j}N_1})$$

for the polygon $\kappa_j=\gamma_0$.
Changing the length of one edge on the arc $A^1R^1B$ on the value
$\delta L$, we have a shift of the point $A^1$ on the distance
$$\delta L\prod_l\cos\alpha_l=\delta L(1+O(\frac{1}{2^{k_j}N_1}))$$

Therefore the variation of length on the second step is small enough,
and a shift of any vertex is no more than $c\delta s$, where $c$ is
some constant independent from $j$. Variation of the magnetic area is
 small enough on the second step.

 Total variation of the functional we get after summation of all
 our contributions:
 $$0=\delta F=(\phi-\int_{R_j}^{Q_j}Bdl+O(\frac{1}{2^{k_j}N_1}))$$

 Lemma is proved.

Proof of the theorem 4. From the lemma 15 above
we have for the limiting curve  $j\rightarrow\infty$
exactly the statement of the theorem. Theorem is proved.

\begin{theo}
For any smooth positive double periodic magnetic field on the
Euclidean plane
(directed along the third axis, orthogonal to the plane) there exist
at least two different periodic convex extremals,
such that the value of the Maupertui-Fermat
functional is positive for them.
\end{theo}

Proof. Let we have only one extremal for the
functional $F$ after the limit $j\rightarrow\infty$.
Consider the overthrowing process
$$f_j:T^2\times [01]\rightarrow
P^a_{2^{k_j}N_1}$$

after the long gradient deformation, determined
by the corrected functional $F^a$. For the large $j$ we have several
(at least
two) different extremals in this space, which have the same limit
for $j\rightarrow\infty$.

Therefore our ''relative extremals'' $\kappa_{j,p},p=1,2,...$ for all
large values $j$
belong to the same
very small contractible neighborhood $W$ of the limiting extremal
$\kappa$ in the space of convex piecewise smooth curves.

After long gradient deformation (mentioned above) the
new overthrowing process belongs to the negative subspace $P^0
\subset P^a_{2^{k_j}N_1}$ everywhere outside of the neighborhood $W$.
As above in the proof of the theorem 2 any overthrowing process of
the point, determined by the map $f_j$, should pass
through this set $W$. By this reason,
the imbedding of the manifold of all one-point curves is contractible
in the space of nonparameterized closed convex curves (i.e.
factor by the action of the group $SO_2$, changing the initial point in the
natural parameterization). But this is an obvious contradiction.
Theorem is proved.

\begin{theo}
Let all periodic convex extremals with the positive value of
Maupertui-Fermat functional are nondegenerate in sense of Morse
in the space of nonparameterized curves.
In this case  there exist at least four periodic convex extremals
for any fixed value of energy such that their Morse indices
are equal to (1,2,2,3).
\end{theo}

For the proof of this theorem we are going to use theorem 3 and
the comparison of Morse indices of periodic extremals with
the Morse indices of  ''relative extremals'' $\kappa_j$ for all values
of $j$ large enough. This comparison (which looks easy),
never was  proved rigorously for our spaces.
So we shall finish the complete proof later. Let us present here
the idea of the proof. A very first question is: {\bf What
is the Morse index for the Maupertui-Fermat functional on the space of
all smooth curves?}.

For the definition of this quantity we have to introduce some unique
receipt of parameterization of curves, because our functional
does not depend on it. A natural parameter (length) is OK for our goals.

After that we consider a Morse index on the space $P$ of the convex
curves with
natural parameterization. This index is finite. There is a trivial ''nullity''
of this critical point equal to 1. It corresponds to the choice
of  initial point on the curve. Our functional is invariant under the
free action of the group $SO_2$ on the space $P$, as it was mentioned above.
The factor-space $\bar{P}$  of the space $P$ by this action
is homotopy equivalent to the torus $T^2$. Our functional
in the generic case has only nondegenerate critical points in $\bar{P}$.

In process of approximation we use the spaces $P_N$ with the natural action
of the group $Z/N\subset SO_2$. After the approximation our functional
is only $Z/N$-invariant. Corresponding factor-spaces $\bar{P}_N$
have the homotopy type $T^2\times S^1$. We found more critical points
in the theorem 3, than we need for the limit $N\rightarrow \infty$.
In fact pairs of them with the neighboring indices $i,i+1$
should have the same limit. Returning to the spaces $P_N$
with discreet parameterization, we get free $Z/N$-critical orbits
instead of the points. This orbits converge to the $SO_2$-orbits
for $N\rightarrow\infty$. Each nondegenerate critical
$SO_2$-orbit with Morse index $i$
generates at least 2 nondegenerate critical $Z/N$-orbits
with Morse indices $i,i+1$
by the  obvious homological reasons {\bf if we shall be able to
prove that this approximation
is really equivalent to the small perturbation  of our functional
in the $C^2$ norm.}

Following the classical papers of Marston Morse, we introduce
a {\bf finite-dimensional approximation} of the space $P$ in the
small neighborhood of the given closed extremal $\gamma$ with
Morse index equal to $i$. It is convenient for us to use the same
spaces $P_N$ of  polygons with $N$ equal edges and total
length not far from the length $NL$ of our extremal $\gamma$.

The approximation of the functional by  Morse is the following:

For $L$ small enough we join all pairs of
the neighboring vertices by the
unique small extremal and construct therefore the {\bf Extremal Polygon}
with the same vertices for any polygon from the space $P_N$.
This space is canonically isomorphic to the space $P_N$,
but the value of our functional
on the extremal polygon is different than its value for the
straight-line polygon with the same vertices. We denote this functional
on the space $P_N$ by $F^e$. {\bf Its extremals are exactly  the same as
 smooth extremals $\gamma$  on the space of all smooth curves.}

 Consider the small neighborhood of the extremal $\gamma$.
 All external angles of polygons in this neighborhood
 have order $O(N^{-1})$. We are going to compare
 functionals $F^e$ and $F$ in this part of the space $P_N$.
 An easy trigonometric
 estimate shows that  difference between these functionals
 has order $O(N^{-2})$.

 More exactly, consider a small straight-line interval $AC$ with length
 equal to $L$
 and a small piece of extremal with the same vertices $A,C$
 (it look like
 an arc of the circle with radius equal to $R_0=B^{-1}$
 in our approximation, for very small
 values of $L$). Here $B$ is a value of the magnetic field in the
 central point of the interval   $AC$.
 Calculating the terms of the order $O(N^{-3})$ for the length of the
 extremal arc and the magnetic
 area of this small domain we shall need also the first derivatives
 of the field $B$ in the same point.After some elementary trigonometry
 we are coming to the following lemma:

 \begin{lem}
 Let our magnetic field belongs to the class
 $C^2$ on the torus.
 The value of  Maupertui-Fermat functional
 for all such ''local'' geometric figures
 bounded by the small straight-line edge and a small piece of the extremal
 is less or equal than the quantity
 $O(L^3)$ with coefficient depending from the maximal values of the
 magnetic field $B$ and its first derivatives on the torus.
 \end{lem}

Combining this result with elementary properties of the  Euclidean Geometry,
we see that any small variation of these local geometric figure
leads to the variation of the functional of the order $O(L^3)$
and $O(L^2\delta L)$ in the variable $N^{-1}$.
If we consider any variation of the polygon from
the small neighborhood of the extremal $\gamma$ with length
$l$, we know a priori that $L\sim O(N^{-1})$ and $\delta L\sim O(N^{-1})$.

 We use now the additivity property of our functional:
 the difference between  functionals $F-F^e$ is equal to
 the sum of $N$ ''local'' terms corresponding to the individual
 edges, described in the lemma above.  Therefore we are coming to
 the following result:

 \begin{lem}
 In the small neighborhood of our extremal $\gamma$ in the space $P_N$
 all derivatives of any order
 of the difference $F-F^e$ are the quantities of the order
 $O(N^{-2})$.
 \end{lem}

Now we are going to finish the proof of the theorem. In the process
of approximation of the space of curves by the extremal polygons
we consider a sequence $N=2^{k_j}N_1$ as above, for which we have
a convergence of the ''relative'' extremals of the functional $F$
 on the subspaces $P_N$
to the smooth extremal $\gamma$. From the old results of Morse we know
that for all large values of $N$ we have the same curve $\gamma$ as
extremal of the
functional $F^e$, i.e. of the same functional on the space of the
extremal polygons.   We know also that
we may consider the tangent spaces to $P_N$ in the point $\gamma$
for all values of $N=2^kN_1$
as the finite dimensional subspaces $T_j$ of the same Hilbert space  $T$.
This sequence $T_j$ converges in the sense that all $T_j$
with larger numbers ''almost '' contain the previous ones:
there exist a natural projector
$$\pi_{j,j+s}:T_j\rightarrow T_{j+s}$$

such that $||\pi_{j,j+s}(u)-u||\rightarrow 0$ for $j\rightarrow \infty$
homogeneously in $s$ and for all unit vectors $u$.
We shall identify the subspaces $T_j$ and
$\pi_{j,j+s}(T_j)$ in our notations.

The second variation of the functional $F^e$ also converges.
It means that this second variation is strictly positive on the subspaces
orthogonal to the image of $T_j$ in $T_{j+s}$ with lowest eigenvalue,
which converges to the $+\infty$ for $j\rightarrow\infty$.
On the image of $T_j$ all lower eigenvalue and eigenvectors
converge to their values on the space of normal vector fields
along the curve $\gamma$. Therefore we may use the finite spaces
$P_N$ for the calculation of the Morse index. Our theorem follows now from
the lemmas above. because the Morse index is stable under the
 perturbations of the function $F^e$ on the spaces $P_N$,
 which are small with first, second
 (and third) derivatives in all points of our neighborhood under
 investigation. The role of 1-dimensional nullity is the following: it
 leads generically to the splitting of one nondegenerate critical circle
 and creation of some nonzero even number of nondegenerate
 critical points in the spaces $P_N$ in the process of approximation:
 half of them with
 index $i$ and another half with index $i+1$; all of them
 converge to our extremal circle which is one point
 $\gamma$ in the space of the
 nonparameterized curves.

Theorem is proved.

\pagebreak

{\bf 4.Appendix.Trigonometric calculations. Proof of the lemmas.}

\vspace{1cm}

Proof of the Lemma 6. It is convenient to decompose the deformation

\noindent
\unitlength=1mm
\begin{picture}(135.00,70.00)
\thicklines
\put(67,3){\makebox(0,0)[cc]{Fig 1.}}
\put(5,10){\vector(1,0){125}}
\put(10,5){\vector(0,1){60}}
\put(130,5){\makebox(0,0)[rb]{x}}
\put(5,65){\makebox(0,0)[lt]{y}}
\put(25,35){\line(2,-1){20}}
\put(45,25){\line(5,-1){22}}
\put(89,25){\line(-5,-1){22}}
\put(109,35){\line(-2,-1){20}}
\put(25,35){\line(3,1){21}}
\put(46,42){\line(4,1){21}}
\put(88,42){\line(-4,1){21}}
\put(109,35){\line(-3,1){21}}
\thinlines
\put(25,35){\vector(1,1){5}}
\put(46,42){\vector(0,1){14}}
\put(67,47.25){\vector(0,1){14}}
\put(88,42){\vector(0,1){14}}
\put(109,35){\vector(-1,1){5}}
\multiput(30,40)(5,-5){3}{\line(1,-1){4}}
\multiput(104,40)(-5,-5){3}{\line(-1,-1){4}}
\multiput(30,40)(5,5){3}{\line(1,1){4}}
\multiput(49,56.75)(6,1.5){3}{\line(4,1){4}}
\multiput(85,56.75)(-6,1.5){3}{\line(-4,1){4}}
\multiput(104,40)(-5,5){3}{\line(-1,1){4}}
\put(24,36){\makebox(0,0)[rb]{$A$}}
\put(29,41){\makebox(0,0)[rb]{$A_t$}}
\put(43,24){\makebox(0,0)[rt]{$E$}}
\put(90,24){\makebox(0,0)[lt]{$F$}}
\put(110,36){\makebox(0,0)[lb]{$B$}}
\put(105,41){\makebox(0,0)[lb]{$B_t$}}
\put(45,42.5){\makebox(0,0)[rb]{$C$}}
\put(89,43){\makebox(0,0)[lb]{$D$}}
\put(45,57){\makebox(0,0)[rb]{$C_t$}}
\put(89,57){\makebox(0,0)[lb]{$D_t$}}
\put(65,55){\makebox(0,0)[lc]{$t$}}
\put(44,49.5){\makebox(0,0)[lc]{$t$}}
\put(86,50){\makebox(0,0)[lc]{$t$}}
\end{picture}

\noindent
in two steps.

\noindent
\unitlength=1mm
\begin{picture}(135.00,35.00)
\put(67,3){\makebox(0,0)[cc]{Fig 2a.}}
\thinlines
\put(10,15){\line(2,-1){10}}
\put(20,10){\line(5,-1){11}}
\put(42,10){\line(-5,-1){11}}
\put(52,15){\line(-2,-1){10}}
\put(10,15){\line(3,1){10.5}}
\put(20.5,18.5){\line(4,1){10.5}}
\put(41.5,18.5){\line(-4,1){10.5}}
\put(52,15){\line(-3,1){10.5}}
\put(9.5,10.5){\makebox(0,0)[rb]{$A$}}
\put(19,9.5){\makebox(0,0)[rt]{$E$}}
\put(42.5,9.5){\makebox(0,0)[lt]{$F$}}
\put(52.5,10.5){\makebox(0,0)[lb]{$B$}}
\put(20,19){\makebox(0,0)[rb]{$C$}}
\put(42,19){\makebox(0,0)[lb]{$D$}}
\put(60,15){\vector(1,0){10}}
\put(85,15){\line(2,-1){10}}
\put(95,10){\line(5,-1){11}}
\put(117,10){\line(-5,-1){11}}
\put(127,15){\line(-2,-1){10}}
\put(85,22){\line(3,1){10.5}}
\put(95.5,25.5){\line(4,1){10.5}}
\put(116.5,25.5){\line(-4,1){10.5}}
\put(127,22){\line(-3,1){10.5}}
\put(85,15){\vector(0,1){7}}
\put(127,15){\vector(0,1){7}}
\put(84.5,10.5){\makebox(0,0)[rb]{$A$}}
\put(94,9.5){\makebox(0,0)[rt]{$E$}}
\put(117.5,9.5){\makebox(0,0)[lt]{$F$}}
\put(127.5,10.5){\makebox(0,0)[lb]{$B$}}
\put(95,27){\makebox(0,0)[rb]{$C_t$}}
\put(117,27){\makebox(0,0)[lb]{$D_t$}}
\put(84.5,23.5){\makebox(0,0)[rb]{$A'$}}
\put(127.5,23.5){\makebox(0,0)[lb]{$B'$}}
\put(83,18.5){\makebox(0,0)[lc]{$t$}}
\put(125,18.5){\makebox(0,0)[lc]{$t$}}
\end{picture}

\begin{center}
\unitlength=0.66mm
\begin{picture}(135.00,80.00)
\put(67,3){\makebox(0,0)[cc]{Fig 2b.}}
\thicklines
\put(10,30){\line(2,-1){40}}
\put(11,58){\line(3,1){41}}
\put(20,40){\line(1,-1){30}}
\put(20,40){\line(1,1){32}}
\put(120,30){\line(-2,-1){40}}
\put(119,58){\line(-3,1){41}}
\put(110,40){\line(-1,-1){30}}
\put(110,40){\line(-1,1){32}}
\thinlines
\put(10,30){\vector(1,1){10}}
\put(120,30){\vector(-1,1){10}}
\put(11,58){\vector(1,-2){9}}
\put(119,58){\vector(-1,-2){9}}
\multiput(10,31.5)(0.25,7){4}{\line(0,1){4}}
\multiput(120,31.5)(-0.25,7){4}{\line(0,1){4}}
\put(8,30){\makebox(0,0)[rc]{$A$}}
\put(8,58){\makebox(0,0)[rc]{$A'$}}
\put(18,41){\makebox(0,0)[rc]{$A_t$}}
\put(122,30){\makebox(0,0)[lc]{$B$}}
\put(122,58){\makebox(0,0)[lc]{$B'$}}
\put(112,41){\makebox(0,0)[lc]{$B_t$}}
\put(52,73){\makebox(0,0)[cb]{$C_t$}}
\put(78,73){\makebox(0,0)[cb]{$D_t$}}
\put(50,8){\makebox(0,0)[ct]{$E$}}
\put(80,8){\makebox(0,0)[ct]{$F$}}
\end{picture}
\unitlength=1mm
\end{center}

Step 1:

We shift the arc $ACDB$ up at the distance $t$. The images of the points
$A$, $B$, $C$, $D$ under the shifts are denoted by $A'$, $B_t$, $C_t$,
$D'$. We also add small vertical segments connecting the points $A$, $A'$
and $B$, $B'$.

Step 2: We rotate the segments $AE$, $A'C_t$, $BF$, $B'D_t$ round the
points $E$, $C_t$, $F$, $D_t$ at such angles that the images of the points
$A$, $A'$ coincide and the images of $B$, $B'$ coincide too. We shall denote
them $A_t$ and $B_t$ respectively.

Let us denote the variation of the magnetic area in the first step and in the
second one by $\delta_1 Q$ and $\delta_2 Q$ respectively. We have
$$
\delta_1 Q \ge B_{min} |AB| t,
$$
$$
\delta_2Q \ge - B_{max} \cdot \frac12 \left(|AE||AA_t|+|C_tA'||A'A_t|+
|BF||BB_t|+|D_tB'||B'B_t| \right ).
$$
The infinitesimal triangles $AA'A_t$ and $BB'B_t$ have the magnetic area
$O(t^2)$ and may be neglected in the first-order calculations.

The angles $AA_tA'$ and $BB_tB'$ are greater then $2\pi/3$ thus
$$
|AA_t|<|AA'|=t, \ |A'A_t|<t, \ |BB_t|<t, \ |B'B_t|<t,
$$
$$
\delta_2Q \ge - B_{max} 2 L t.
$$
$$
|AB| \ge \frac{LN}4 \ge \frac{LN_0}{4}> 2 \frac{B_{max}}{B_{min}} L
$$
Combining all this estimates we get:
$$
\delta Q =\delta_1 Q +\delta_2 Q > 2 B_{max} L t -2 B_{max} L t =0.
$$
The Lemma 6 is proved.

\eject

Proof of the Lemma 8. It is convenient to decompose the deformation
in two steps.

Step 1:

\unitlength=1mm
\noindent
\begin{picture}(135.00,85.00)
\thicklines
\put(67,3){\makebox(0,0)[cc]{Fig 3.}}
\put(20,50){\line(5,2){25}}
\put(45,59.75){\line(5,1){25}}
\put(95,58.5){\line(-4,1){25}}
\put(120,50){\line(-3,1){25}}
\put(20,50){\line(1,-1){20}}
\put(40,30){\line(5,-2){25}}
\put(90,20){\line(-1,0){25}}
\put(115,25){\line(-5,-1){25}}
\put(120,50){\line(-1,-5){5}}
\thinlines
\multiput(20,50)(2.5,0.74){21}{\circle*{0.2}}
\multiput(120,50)(-2.5,0.74){21}{\circle*{0.2}}
\multiput(20,50)(2.25,-1.5){21}{\circle*{0.2}}
\multiput(120,50)(-2.75,-1.5){21}{\circle*{0.2}}
\put(5,50){\vector(1,0){125}}
\put(10,5){\vector(0,1){75}}
\put(130,45){\makebox(0,0)[rb]{x}}
\put(5,80){\makebox(0,0)[lt]{y}}
\multiput(70,65)(0,-5.5){3}{\line(0,-1){4}}
\multiput(65,20)(0,6.5){5}{\line(0,1){4}}
\put(70,65){\vector(-1,3){3.33}}
\put(70,65){\vector(1,3){3.33}}
\multiput(68.33,75)(1.66,0){3}{\circle*{0.2}}
\put(65,20){\vector(-2,-3){3.33}}
\put(65,20){\vector(2,-3){3.33}}
\multiput(63.33,15)(1.66,0){3}{\circle*{0.2}}
\multiput(20,50)(2.33,1.25){21}{\circle*{0.2}}
\multiput(120,50)(-2.33,1.25){21}{\circle*{0.2}}
\multiput(20,50)(2.08,-1.75){21}{\circle*{0.2}}
\multiput(120,50)(-2.58,-1.75){21}{\circle*{0.2}}
\multiput(20,50)(1.166,0.75){21}{\circle*{0.2}}
\multiput(43.33,65)(1.166,0.5){21}{\circle*{0.2}}
\multiput(20,50)(0.941,-1.125){21}{\circle*{0.2}}
\multiput(38.83,27.5)(1.14,-0.625){21}{\circle*{0.2}}
\multiput(120,50)(-1.166,0.675){21}{\circle*{0.2}}
\multiput(96.66,63.5)(-1.166,0.575){21}{\circle*{0.2}}
\multiput(120,50)(-0.125,-1.26){21}{\circle*{0.2}}
\multiput(117.5,24.8)(-1.225,-0.39){21}{\circle*{0.2}}
\multiput(93,17)(-1.233,-0.1){21}{\circle*{0.2}}
\put(30,51){\makebox(0,0)[lb]{$\varphi_1$}}
\put(110,51){\makebox(0,0)[rb]{$\varphi_2$}}
\put(110,49){\makebox(0,0)[rt]{$\varphi_3$}}
\put(30,49){\makebox(0,0)[lt]{$\varphi_4$}}
\put(19,51){\makebox(0,0)[rb]{$A$}}
\put(121,51){\makebox(0,0)[lb]{$B$}}
\put(71,63){\makebox(0,0)[lt]{$C$}}
\put(64,23.5){\makebox(0,0)[rb]{$D$}}
\put(71,51){\makebox(0,0)[lb]{$\tilde C$}}
\put(64,49){\makebox(0,0)[rt]{$\tilde D$}}
\put(66.6,76){\makebox(0,0)[rb]{$C_t$}}
\put(73.3,76){\makebox(0,0)[lb]{$C'$}}
\put(61.6,14){\makebox(0,0)[rt]{$D_t$}}
\put(68.6,14){\makebox(0,0)[lt]{$D'$}}
\end{picture}

We rotate the whole arcs $AC$, $BC$, $AD$, $BD$ around the points $A$ and
$B$ to the corresponding angles
\begin{eqnarray*}
\nonumber
\delta \varphi_1=\frac{1}{\cos \varphi_1} \cdot \frac{1}{\tan \varphi_1+
\tan \varphi_2} \cdot \frac{\delta x}{|AC|}, \
\delta \varphi_2=\frac{1}{\cos \varphi_2} \cdot \frac{1}{\tan \varphi_1+
\tan \varphi_2} \cdot \frac{\delta x}{|BC|}
\\
\delta \varphi_3=\frac{1}{\cos \varphi_3} \cdot \frac{1}{\tan \varphi_3+
\tan \varphi_4} \cdot \frac{\delta x}{|BD|}, \
\delta \varphi_4=\frac{1}{\cos \varphi_4} \cdot \frac{1}{\tan \varphi_3+
\tan \varphi_4} \cdot \frac{\delta x}{|AD|}
\end{eqnarray*}
Here $\varphi_1$, $\varphi_2$, $\varphi_3$, $\varphi_4$ are the angles
between the $x$-axes and the intervals $AC$, $BC$, $BD$, $AD$ respectively,
$|AC|$ denotes the length of the span $AC$.

Let us denote the images of the points $C$, $D$ after the rotations around
the point $A$ by $C_t$, $D_t$, the images of the points $C$, $D$ after
the rotations around the point $B$ by $C'$, $D'$, the ortohonal projections
of the points $C$ and $D$ to the interval $AB$ by $\tilde C$, $\tilde D$.

Finally we add small horizontal intervals connecting $C_t$ and $C'$, $D_t$
and $D'$.

Step 2:
\nopagebreak

\unitlength=1mm
\noindent
\begin{picture}(135.00,85.00)
\thicklines
\put(67,3){\makebox(0,0)[cc]{Fig 4.}}
\put(20,50){\line(5,3){23.33}}
\put(43.33,64){\line(5,2){25}}
\put(74.66,74){\line(5,-2){25}}
\put(99.66,64){\line(5,-3){23.33}}
\put(69.66,15){\line(6,1){24.33}}
\put(94,19.06){\line(4,1){24}}
\put(118,25.06){\line(1,5){4.99}}
\put(20,50){\line(4,-5){18}}
\put(38,27.5){\line(2,-1){25}}
\thinlines
\multiput(68.33,74)(6.89,-2.76){4}{\line(5,-2){4}}
\multiput(93,64)(6.41,-3.85){4}{\line(5,-3){4}}
\multiput(63,15)(6.66,1.11){4}{\line(6,1){4}}
\multiput(87.33,19.06)(7,1.75){4}{\line(4,1){4}}
\multiput(111.33,25.06)(1.33,6.65){4}{\line(1,5){1}}
\put(69.66,15){\vector(-1,0){6.66}}
\put(94,19.06){\vector(-1,0){6.66}}
\put(118,25.06){\vector(-1,0){6.66}}
\put(123,50){\vector(-1,0){6.66}}
\put(74.66,74){\vector(-1,0){6.33}}
\put(99.66,64){\vector(-1,0){6.66}}
\put(19,51){\makebox(0,0)[rb]{$A$}}
\put(124,51){\makebox(0,0)[lb]{$B$}}
\put(115,49){\makebox(0,0)[rc]{$B_t$}}
\put(68.33,75){\makebox(0,0)[rb]{$C_t$}}
\put(75,75){\makebox(0,0)[lb]{$C'$}}
\put(63,14){\makebox(0,0)[rt]{$D_t$}}
\put(69.66,14){\makebox(0,0)[lt]{$D'$}}
\put(114.5,26){\makebox(0,0)[cb]{$\delta x$}}
\put(119.5,49){\makebox(0,0)[ct]{$\delta x$}}
\end{picture}

We shift the whole arc $C'BD'$ left at the distance $\delta x$.

Let us denote the magnetic area above the line $AB$ and below the line $AB$
by $Q_+$ and $Q_-$ respectively, $Q=Q_+ +Q_-$, $Q_{AD}$ be the magnetic
area of the polygon bounded by the arc $AD$ and by the span $AD$,
$Q_{BD}$ be the magnetic area of the polygon bounded by the arc $BD$ and
by the span $BD$, $\delta_1$ be the variation in the first step,
$\delta_2$ be the variation in the second step. Then

\begin{equation}
\label{l8-c2}
\delta_1 Q_+ \ge \frac12 B_{min} |AC||CC_t| +\frac12 B_{min} |BC||CC'|
\end{equation}
\begin{eqnarray}
\nonumber
\delta_1 Q_- \ge \delta_1 Q_{AD} + \delta_1 Q_{BD} +
B_{min} \cdot \hbox{area of the triangle } ADD_t  +
\\ \label{l8-c3}
+B_{min} \cdot \hbox{area of the triangle } BDD'
\end{eqnarray}
\begin{equation}
\label{l8-c4}
\delta_2 Q \ge - \frac{NL}2 B_{max} \delta x
\end{equation}

We shall use the following estimates:
\begin{description}
\item[1)] $\varphi_1+\varphi_2 < N\alpha_0/2$, $\varphi_1< N\alpha_0/2$,
$\varphi_2 < N\alpha_0/2$, $\cos \varphi_1 >0.9$, $\cos \varphi_2 >0.9$,
$$\tan \varphi_1 < \frac{9 B_{min}}{20 B_{max} (2N^3+N/2)},\
\tan \varphi_2 < \frac{9 B_{min}}{20 B_{max} (2N^3+N/2)}.$$
\item[2)] The angles between the $x$-axis and all the segments of the
arc $AB$ are less then $N \alpha_0 /2$, their cosines are greater then
0.9.
\item[3)] $|AB| >1.8L$, $|A\tilde C|>0.4L$, $|A\tilde D|>0.4L$,
$|B\tilde C|>0.4L$, $|B\tilde D|>0.4L$, $|AC|>0.4L$, $|AD|>0.4L$,
$|BC|>0.4L$, $|BD|>0.4L$.
\item[4)] $|D\tilde D| > L/2N$.
\end{description}
The estimates 1) - 3) follow directly from the definition of $\alpha_0$.
Let us prove the estimate 4).

Due to the Lemma 6 it follows that if $\gamma$ is an extremal then at
least on of the arcs $AD$ or $BD$ has no internal angles less then
$\pi/3$. For the sake of concreteness let us assume that the arc $AD$
has this property. Then there are two possibilities.

1) Moving from the point $A$ to the point $D$ along the arc $AD$ we always
move right. Let us denote $Q$ the neighboring vertex to $A$ in the arc $AD$.
The angle between the $x$-axes and $AQ$ is greater then $\pi/3-N\alpha_0/2 >
\pi/3-0.001$.

2) Moving from the point $A$ to the point $D$ along the arc $AD$ we move
left and then right. Let $T$ be turning vertex, $P$ and $Q$ be the preceding
and the succeeding ones. Then the projection of the interval $PQ$ to the
$y$-axes is greater then $\sqrt 3 L/2$.

\unitlength=1mm
\noindent
\begin{picture}(135.00,50.00)
\thicklines
\put(67,3){\makebox(0,0)[cc]{Fig 5.}}
\put(70,40){\line(-4,-1){20}}
\put(50,35){\line(-2,-1){20}}
\put(30,25){\line(1,-1){15}}
\put(45,10){\line(1,0){20}}
\put(65,10){\line(4,1){20}}
\put(85,15){\line(3,1){20}}
\put(105,21.66){\line(3,2){20}}
\put(125,35){\line(-4,1){20}}
\thinlines
\put(5,40){\vector(1,0){125}}
\put(10,5){\vector(0,1){40}}
\put(130,35){\makebox(0,0)[rb]{x}}
\put(5,45){\makebox(0,0)[lt]{y}}
\put(45,10){\line(2,1){60}}
\put(85,15){\line(0,1){25}}
\put(70,41){\makebox(0,0)[cb]{$A$}}
\put(105,41){\makebox(0,0)[cb]{$B$}}
\put(49,36){\makebox(0,0)[rb]{$P$}}
\put(29,26){\makebox(0,0)[rb]{$T$}}
\put(44,9){\makebox(0,0)[rt]{$Q$}}
\put(86,14){\makebox(0,0)[lt]{$D$}}
\put(86,41){\makebox(0,0)[lb]{$\tilde D$}}
\put(86,29){\makebox(0,0)[lt]{$S$}}
\put(130,35){\makebox(0,0)[rb]{x}}
\end{picture}

In the both cases the distance between $Q$ and the line $AB$ is greater then
$(\sqrt3/2-0.001) L$. Let us connect $Q$ with $B$ and denote the crossing point
of the intervals $QB$ and $D\tilde D$ by $S$. $|QB|<NL/2$, $|SB|>0.4$ then
$$
|D\tilde D| > |SD| > \frac{|SB|}{|QB|} \left (\frac{\sqrt3}2-0.001  \right )L
> \frac{0.4L}{NL/2} \left (\frac{\sqrt3}2-0.001  \right )L >\frac L{2N}
$$

For the infinitesimal intervals $CC_t$, $CC'$ we have:
$$
|CC_t|=|AC|\delta \varphi_1=\frac{1}{\cos \varphi_1} \cdot
\frac{1}{\tan \varphi_1+ \tan \varphi_2} \delta x >
\frac{1}{\tan \varphi_1+ \tan \varphi_2} \delta x,
$$
thus
$$
|CC_t|>\frac{10 B_{max}(2N^3+N/2)}{9 B_{min} } \delta x, \
|CC'|>\frac{10 B_{max}(2N^3+N/2)}{9 B_{min} } \delta x.
$$
$|AC|+|BC|>|A\tilde C|+|B\tilde C| >1.8L, \hbox{ thus}$
\begin{equation}
\label{l8-c5}
\delta_1Q_+>0.9B_{min} L \frac{10 B_{max}(2N^3+N/2)}{9 B_{min} } \delta x
=(2N^3+\frac N2)LB_{max} \delta x.
\end{equation}
$$
\delta \varphi_3=\frac{1}{\cos \varphi_3} \cdot \frac{1}{\tan \varphi_3+
\tan \varphi_4} \cdot \frac{\delta x}{|BD|}<
\frac{1}{\cos \varphi_3} \cdot \frac{1}{\tan \varphi_3}
\cdot \frac{\delta x}{|BD|} =
\frac{1}{\sin \varphi_3} \cdot \frac{\delta x}{|BD|}
$$
$\sin\varphi_3 |BD|=|D\tilde D|$ thus
$$
\delta \varphi_3< \frac{2N}{L} \delta x, \
\delta \varphi_4< \frac{2N}{L} \delta x.
$$

Applying the Lemma 5 and taking into account that the magnetic areas
of the triangles $ADD_t$ and $BDD'$ are positive we get
\begin{equation}
\label{l8-c6}
\delta_1Q_- \ge -2N^3 L B_{max} \delta x.
\end{equation}

Combining (\ref{l8-c4}), (\ref{l8-c5}),(\ref{l8-c6}) we get:
$$
\delta Q=\delta_1 Q_+ +\delta_1 Q_- + \delta_2 Q >0.
$$

This completes the proof.

\eject

Proof of the Lemma 12. To estimate the variation of the magnetic area $Q$
under this deformation it is convenient to decompose this deformation
in two steps.

Step 1:

\noindent
\unitlength=1mm
\begin{picture}(110.00,85.00)
\thicklines
\put(67,3){\makebox(0,0)[cc]{Fig 6.}}
\put(10,45){\line(1,4){4}}
\put(14,61){\line(5,2){15}}
\put(29,67){\line(5,1){16}}
\put(45,70.2){\line(4,-1){16}}
\put(61,66.2){\line(5,-2){15}}
\put(76,60.2){\line(2,-1){15}}
\put(10,45){\line(1,-3){5}}
\put(15,30){\line(2,-1){15}}
\put(30,22.5){\line(4,-1){16}}
\put(46,18.5){\line(1,0){16}}
\put(62,18.5){\line(2,1){15}}
\put(77,26){\line(3,2){14}}
\put(91,35.33){\line(0,1){17.37}}
\thinlines
\put(14,61){\line(1,0){20}}
\put(15,30){\line(1,0){20}}
\put(91,52.7){\line(-1,0){20}}
\put(91,35.33){\line(-1,0){20}}
\put(26,58){\vector(-1,1){5}}
\put(28,33){\vector(-1,-1){5}}
\put(79,50){\vector(1,1){5}}
\put(79,38){\vector(1,-1){5}}
\put(27,57){\makebox(0,0)[lt]{$\varphi_1$}}
\put(78,49){\makebox(0,0)[rt]{$\varphi_2$}}
\put(78,39){\makebox(0,0)[rb]{$\varphi_3$}}
\put(29,34){\makebox(0,0)[lb]{$\varphi_4$}}
\multiput(29,74)(6,1.2){3}{\line(5,1){4}}
\multiput(45,77.2)(6,-1.5){3}{\line(4,-1){4}}
\multiput(61,73.2)(5.5,-2.2){3}{\line(5,-2){4}}
\put(29,67){\vector(0,1){7}}
\put(45,70.2){\vector(0,1){7}}
\put(44,73){\makebox(0,0)[rc]{$\delta y_1$}}
\put(61,66.2){\vector(0,1){7}}
\put(60,70.7){\makebox(0,0)[rc]{$\delta y_1$}}
\put(76,60.2){\vector(0,1){7}}
\put(29,67){\vector(-2,3){4.66}}
\multiput(25.88,74)(1.53,0){2}{\circle*{0.2}}
\put(76,60.2){\vector(1,1){7}}
\multiput(76.87,67.2)(1.75,0){3}{\circle*{0.2}}
\multiput(15,62.5)(5,6.25){2}{\line(4,5){3.5}}
\multiput(91,52.7)(-4.5,8){2}{\line(-3,5){3}}
\multiput(30,16.5)(6,-1.5){3}{\line(4,-1){4}}
\multiput(46,12.5)(6,0){3}{\line(1,0){4}}
\multiput(62,12.5)(5.5,2.75){3}{\line(2,1){4}}
\put(30,22.5){\vector(0,-1){6}}
\put(46,18.5){\vector(0,-1){6}}
\put(45,16.5){\makebox(0,0)[rc]{$\delta y_2$}}
\put(62,18.5){\vector(0,-1){6}}
\put(61,15.5){\makebox(0,0)[rc]{$\delta y_2$}}
\put(77,26){\vector(0,-1){6}}
\put(30,22.5){\vector(-1,-1){6}}
\multiput(25.5,16.5)(1.5,0){3}{\circle*{0.2}}
\put(77,26){\vector(1,-1){6}}
\multiput(78.5,20)(1.5,0){3}{\circle*{0.2}}
\multiput(15,30)(5,-7.5){2}{\line(2,-3){4}}
\multiput(91,35.33)(-5,-9.33){2}{\line(-1,-2){3}}
\put(15,60){\makebox(0,0)[lt]{$A$}}
\put(30,66){\makebox(0,0)[lt]{$B$}}
\put(23.5,75){\makebox(0,0)[rb]{$B_t$}}
\put(30,76){\makebox(0,0)[lb]{$B'$}}
\put(75,59.2){\makebox(0,0)[rt]{$C$}}
\put(76,70.2){\makebox(0,0)[rb]{$C'$}}
\put(84,68.2){\makebox(0,0)[lb]{$C''$}}
\put(90,51.7){\makebox(0,0)[rt]{$D$}}
\put(90,36.33){\makebox(0,0)[rb]{$E$}}
\put(76,27){\makebox(0,0)[rb]{$F$}}
\put(76,17){\makebox(0,0)[rt]{$F'$}}
\put(84,19){\makebox(0,0)[lt]{$F''$}}
\put(31,23.5){\makebox(0,0)[lb]{$G$}}
\put(31,14.5){\makebox(0,0)[lt]{$G'$}}
\put(23,15.5){\makebox(0,0)[rt]{$G_t$}}
\put(16,31){\makebox(0,0)[lb]{$H$}}
\end{picture}

We shift the arc $BC$ up at the distance $\delta y_1$, the arc $FG$
down at the distance $\delta y_2$ and we rotate the segments $AB$, $CD$,
$EF$, $GH$ over the points $A$, $D$, $E$, $H$ at the angles
$$
\delta \varphi_1=\frac{\delta y_1}{L\cos \varphi_1}, \
\delta \varphi_2=\frac{\delta y_1}{L\cos \varphi_2}, \
\delta \varphi_3=\frac{\delta y_2}{L\cos \varphi_3}, \
\delta \varphi_4=\frac{\delta y_2}{L\cos \varphi_4},
$$
where  $\varphi_1$, $\varphi_2$, $\varphi_3$, $\varphi_4$ are the angles
between the real line and the segments $AB$, $CD$, $EF$, $GH$,
$|\tan \varphi_k| \le 1$, $k=1,\ldots,4$. Let us denote the images of the
points $B$, $C$, $F$, $G$ under the shifts $B'$, $C'$, $F'$, $G'$, the
images under the rotations $B_t$, $C''$, $F''$, $G_t$.

Step 2:
\nopagebreak

\noindent
\begin{picture}(110.00,85.00)
\thicklines
\put(67,3){\makebox(0,0)[cc]{Fig 7.}}
\put(10,45){\line(1,4){4}}
\put(14,61){\line(4,5){10.4}}
\put(29,74){\line(5,1){16}}
\put(45,77.2){\line(4,-1){16}}
\put(61,73.2){\line(5,-2){15}}
\put(83,67.2){\line(3,-5){8}}
\put(10,45){\line(1,-3){5}}
\put(15,30){\line(2,-3){9}}
\put(30,16.5){\line(4,-1){16}}
\put(46,12.5){\line(1,0){16.4}}
\put(62.4,12.5){\line(2,1){15}}
\put(83,20){\line(1,2){8}}
\put(91,36){\line(0,1){17.86}}
\thinlines
\multiput(24.4,74)(6,1.2){3}{\line(5,1){4}}
\multiput(40.4,77.2)(6,-1.5){3}{\line(4,-1){4}}
\multiput(56.4,73.2)(5.5,-2.2){3}{\line(5,-2){4}}
\put(29,74){\vector(-1,0){4.6}}
\put(45,77.2){\vector(-1,0){4.6}}
\put(43,79){\makebox(0,0)[cb]{$\delta x_1$}}
\put(61,73.2){\vector(-1,0){4.6}}
\put(59,75){\makebox(0,0)[cb]{$\delta x_1$}}
\put(83,67.2){\vector(-1,0){11.6}}
\put(91,53.86){\vector(-1,0){11.6}}
\put(84,54.8){\makebox(0,0)[cb]{$\delta x$}}
\put(91,36){\vector(-1,0){11.6}}
\put(84,35){\makebox(0,0)[ct]{$\delta x$}}
\multiput(24,16.5)(6,-1.5){3}{\line(4,-1){4}}
\multiput(56.4,12.5)(5.5,2.75){3}{\line(2,1){4}}
\put(30,16.5){\vector(-1,0){6}}
\put(46,12.5){\vector(-1,0){6}}
\put(43,11.5){\makebox(0,0)[ct]{$\delta x_2$}}
\put(62,12.5){\vector(-1,0){6}}
\put(59,11.5){\makebox(0,0)[ct]{$\delta x_2$}}
\put(83,20){\vector(-1,0){11.6}}
\multiput(71.4,67.2)(3,-5){3}{\line(3,-5){2.2}}
\multiput(71.4,20)(3,6){3}{\line(1,2){2}}
\multiput(79.4,36)(0,6.5){3}{\line(0,1){4}}
\put(15,60){\makebox(0,0)[lt]{$A$}}
\put(23.5,75){\makebox(0,0)[rb]{$B_t$}}
\put(30,73){\makebox(0,0)[lt]{$B'$}}
\put(77,70.2){\makebox(0,0)[cb]{$C'$}}
\put(84,68.2){\makebox(0,0)[lb]{$C''$}}
\put(70.4,66.2){\makebox(0,0)[rt]{$C_t$}}
\put(90,52.8){\makebox(0,0)[rt]{$D$}}
\put(90,37){\makebox(0,0)[rb]{$E$}}
\put(78.4,52.8){\makebox(0,0)[rt]{$D_t$}}
\put(78.4,37){\makebox(0,0)[rb]{$E_t$}}
\put(77,17){\makebox(0,0)[ct]{$F'$}}
\put(84,19){\makebox(0,0)[lt]{$F''$}}
\put(70.4,21){\makebox(0,0)[rb]{$F_t$}}
\put(31,17.5){\makebox(0,0)[lb]{$G'$}}
\put(23,15.5){\makebox(0,0)[rt]{$G_t$}}
\put(16,31){\makebox(0,0)[lb]{$H$}}
\end{picture}

We shift the arc $B'C'$ left at the distance $\delta x_1=L \sin \varphi_1
\delta \varphi_1=\tan \varphi_1 \ \delta y_1$ the arc $F'G'$ left at the
distance $\delta x_2=L \sin \varphi_4 \delta \varphi_4=\tan \varphi_4 \ \delta
y_2$, the arc $C''DEF''$ left at the distance
\begin{equation}
\label{l12-c1}
\delta x= \delta y_1(\tan \varphi_1 \ +\tan \varphi_2 \ )=
\delta y_2(\tan \varphi_3 \ + \tan \varphi_4 \ ).
\end{equation}
{}From (\ref{l12-c1}) it follows that $\delta y_1=t/(\tan \varphi_1 \ +\tan
\varphi_2 \ )$, $\delta y_2=t/(\tan \varphi_3 \ + \tan \varphi_4 \ )$.

Let us denote $BC_x$ and $FG_x$ the lengths of the projections of the
arcs $BC$ and $FG$ to the x-axes.

For the change of magnetic area in the step 1 we have
$$
\delta_1 Q \ge B_{min} (\delta y_1 BC_x + \delta y_2 FG_x) =
B_{min} \left [ \frac{t \, BC_x}{\tan \varphi_1 + \tan \varphi_2} +
\frac{t \, FG_x}{\tan \varphi_3 + \tan \varphi_4} \right ]
$$
thus
$$
\delta_1 Q \ge B_{min} t \min(BC_x,FG_x)
$$

The angles between the $x$-axes and all segments of the arcs $AH$ and $DE$
are greater then $\pi/4$ thus the projection of these arcs to the $x$-axes
are smaller then $D_{min}$ and
$$
\min(BC_x,FG_x) \ge D_x-2D_{min}-2L,
$$
where $D_x$ denotes the diameter of $\gamma$ in the direction $x$.  It
is easy to show that
$$
D_x \ge D_{max}-D_{min}, \ L \le \frac4N D_{max}.
$$
Thus
$$
\delta_1 Q \ge B_{min} t \left [ \left (1-\frac8N \right ) D_{max} -
3D_{min} \right ]
$$

For the variation of the magnetic area in the second step we have
$$
\delta_2 Q \ge -B_{max} D_{min} t.
$$
Thus
$$
\delta Q = \delta_1 Q + \delta_2 Q \ge
\left \{B_{min}\left [ \left (1-\frac8N \right ) D_{max} -
3D_{min} \right ] - B_{max} D_{min}\right \} t.
$$
If
$$
\frac{D_{max}}{D_{min}} > \frac{B_{max} B_{min}^{-1}+3}{1-8N^{-1}}
$$
then
$$
\delta Q \ge
\left \{B_{min}\left [ B_{max} B_{min}^{-1} D_{max} +3D_{min} -
3D_{min} \right ] - B_{max} D_{min}\right \} t >0.
$$
Lemma 12 is proved.

Proof of the Lemma 13.

\begin{center}
\unitlength=1mm
\noindent
\begin{picture}(60.00,40.00)
\thicklines
\put(30,1){\makebox(0,0)[cb]{Fig 8.}}
\put(10,20){\line(4,3){14}}
\put(24,30.5){\line(1,0){16.5}}
\put(40.5,30.5){\line(4,-3){14}}
\put(10,20){\line(1,-1){12}}
\put(22,8){\line(4,1){17}}
\put(39,12.25){\line(2,1){15.5}}
\put(24,30.5){\vector(-4,-3){4}}
\put(22,8){\vector(-1,1){4}}
\put(40.5,30.5){\vector(-1,0){9}}
\put(39,12.25){\vector(-4,-1){9}}
\thinlines
\multiput(21,27.66)(6,1.5){2}{\line(4,1){4}}
\multiput(31.9,29.7)(2.1,-5.3){2}{\line(2,-5){1.6}}
\multiput(19,11.83)(6,-1){2}{\line(6,-1){4}}
\multiput(30,10)(3,4.5){2}{\line(2,3){2.4}}
\put(36,19){\circle*{0.2}}
\multiput(33,30.05)(2,-0.901){10}{\circle*{0.2}}
\multiput(31.25,10.416)(2,0.833){11}{\circle*{0.2}}

\put(9,21){\makebox(0,0)[rb]{$B$}}
\put(55.5,21){\makebox(0,0)[lb]{$A$}}
\put(36,20){\makebox(0,0)[lb]{$A_t$}}
\put(24,31.5){\makebox(0,0)[rb]{$C$}}
\put(19,28.5){\makebox(0,0)[rb]{$C_t$}}
\put(22,7){\makebox(0,0)[rt]{$D$}}
\put(17,11){\makebox(0,0)[rt]{$D_t$}}
\put(31.5,31.5){\makebox(0,0)[cb]{$P_t$}}
\put(41,31){\makebox(0,0)[lb]{$P$}}
\put(40,11.25){\makebox(0,0)[lt]{$Q$}}
\put(30,9){\makebox(0,0)[t]{$Q_t$}}
\end{picture}
\end{center}
The variation of the magnetic area under this deformation consists of
two parts:

1) The magnetic area of the small triangles near all
the vertices except $A$ and $B$ (we shall denote it by $\delta_1 Q$).

2) The magnetic area of the small quadrangle $AP_tA_tQ_t$
near the vertex $A$. Here we denote the neighbors of $A$ by $P$ and $Q$,
the shifts of the points $A$, $P$, $Q$ by $A_t$, $P_t$, $Q_t$,
the magnetic area of the quadrangle by $\delta_2 Q$.

We use the following estimate. Let $F$ be a vertex of our polygon,
the arc $BF$ contains $k$ segments. Then the shift of the vertex $F$
under the deformation $FF_t$ is less then $kt$. It is easy to prove it
by induction.

\begin{center}
\unitlength=1mm
\noindent
\begin{picture}(60.00,30.00)
\put(30,1){\makebox(0,0)[cb]{Fig 9.}}
\thicklines
\put(30,20){\line(1,0){25}}
\put(30,20){\line(-2,-1){20}}
\put(55,20){\vector(-1,0){10}}
\put(30,20){\vector(-2,-1){10}}
\thinlines
\multiput(30,20)(2,1){7}{\circle*{0.2}}
\multiput(20.5,15.1)(5,1){5}{\line(5,1){4}}
\put(37,21){\makebox(0,0)[lb]{$\alpha_E$}}
\put(30,21){\makebox(0,0)[rb]{$E$}}
\put(20,16){\makebox(0,0)[rb]{$E_t$}}
\put(55,21){\makebox(0,0)[cb]{$F$}}
\put(46,22){\makebox(0,0)[cb]{$F_t$}}
\end{picture}
\end{center}

Let $E$ be the neighbor of $F$ on the arc $BF$, $\alpha_E$ be the external
angle in the vertex $E$, $E_t$ be the shift of $E$. Then
$$
|FF_t|=|EE_t| \cos \alpha_E \, + t < |EE_t|+t,
$$
Then the shifts of all the vertices except $A$ are less then $Nt/2$. For
the area of the triangle near the vertex $E$ we have
$$
\hbox{area of the triangle}\, E_tEF_t =\frac12|EE_t||EF_t| \sin \alpha_E <
\frac12 NLt \alpha_E< \frac{{\cal L}_0}{2}t \alpha_E.
$$
The sum of all external angles is equal to $2\pi$ thus the sum of the areas
of all triangles is less then $\pi {\cal L}_0t$ and the magnetic area of these
triangles is less then $B_{max}\pi {\cal L}_0t$.

The distances $|AA_t|$ and $|P_tQ_t|$ can be estimated by
$$
|AA_t| \le 2 \frac{Nt}{2} \frac{1}{\sin \alpha}, \
|P_tQ_t| < 2 \frac{{\cal L}_0}N,
$$
where $\alpha$ is the external angle in the vertex $A$.
Then for the area of the quadrangle $AP_tA_tQ_t$ we have
$$
\hbox{area of the quadrangle}\ AP_tA_tQ_t \le \frac12|P_tQ_t||AA_t| <
{\cal L}_0 t (\sin \alpha)^{-1}.
$$
Finally we get
$$
\delta F=\delta l(\gamma) -\delta Q < -N t + B_{max}\pi {\cal L}_0t
+ {\cal L}_0 B_{max} t(\sin \alpha)^{-1}.
$$

\pagebreak

\vspace{1cm}

\pagebreak

\end{document}